\documentclass{ws-procs9x6}
\usepackage{amsfonts}
\usepackage{amssymb}
\def\be{\begin{eqnarray}}
\def\ee{\end{eqnarray}}

\begin{document}

\title{Branes, Strings, and Odd \\  Quantum Nambu Brackets} 

\author{T L CURTRIGHT} 

\address{Department of Physics, University of Miami,\\
Coral Gables, FL 33124-8046, USA\\ 
E-mail: curtright@physics.miami.edu} 

\author{C K ZACHOS}

\address{High Energy Physics Division,\\ 
Argonne National Laboratory,\\
Argonne, IL 60439-4815, USA \\
E-mail: zachos@anl.gov}  

\maketitle

\abstracts{The dynamics of topological open branes is controlled by 
Nambu Brackets. Thus, they might be quantized through the 
consistent quantization of the underlying Nambu brackets, 
including odd ones: these are reachable systematically from even brackets, 
whose more tractable properties have been detailed before. }

\section{Classical Nambu Dynamics}\label{sec:classical} 

The classical motion of topological open membranes is controlled by Nambu 
Brackets, the multilinear generalization of Poisson Brackets.\cite{nambu} 
We have addressed even branes before, in Refs \refcite{sphere,CQNB,talks}, but 
not odd ones, treated here, essentialy by reduction to even ones, in a 
parallel solution of the problem.

Consider the ``vortex string"  2-form 
action,\cite{estabrook,rasetti,lund,takhtajan,talks}
\be
S=\int \Bigl ( z^1 ~ d z^2 \wedge dz^3 - H ~ dG \wedge  dt\Bigr ) ~,
\ee
originating in an exact 3-form, 
\be 
d\omega_2 = dz^1 \wedge dz^2 \wedge dz^3 - 
dH \wedge  dG \wedge  dt~,
\label{3form}
\ee 
(in analogy to the Hamilton-Poincar\'{e} symplectic 
2-form, $d\omega_1=dx \wedge dp_x + dH \wedge  dt$). 
The 3-integral of this form on an open 3-surface yields 
the above 2-form action evaluated on the 2-boundary of that surface,
{\em i.e.}, a Cartan integral invariant, analogous to the $(1+1)$-dimensional 
$\sigma$-model  WZW topological interaction terms or Chern-Simons actions.  

More explicitly, in world-sheet coordinates $t, \sigma$, 
the action reads 
\be
S=\int dt d\sigma ~ \left ( {\epsilon^{ijk}\over 3}
 z^i \partial_t z^j \partial_\sigma z^k + H\partial_\sigma G \right ) ~ .
\label{vortexstringaction} 
\ee
For such systems, Dirac quantization procedures\footnote{ 
The Lagrangean in this action is singular,\cite{rasetti} 
since it is linear in the velocities,
yielding three primary constraints $\phi^i= \pi^i +{\epsilon^{ijk}\over 3}
 z^j \partial_\sigma z^k=0$, involving the canonical momentum
densities $\pi^i(\sigma)$ conjugate to the $z^i$s. 
Taking Poisson brackets (PB) of these constraints
with the Hamiltonian (which is minus the second term in the Lagrangean) 
yields further, secondary, constraints,\cite{Dirac} 
{\em etc.},  in an arduous iterative procedure.
The mutual PBs for some of these constraints vanish, but not for 
some others: the ensuing second class constraints could be projected 
out through Dirac brackets (DB), which specify novel (non-flat) 
Poisson structures, which may then lead to quantization.\cite{Dirac,rasetti} }
have been outlined in, {\em e.g.}, 
Refs \refcite{rasetti,mukunda,Bayen}; but they are cumbersome to complete, 
insofar as they rely on Kontsevich's  generalization\cite{Kontsevich} 
of Moyal Brackets to nonflat Poisson manifolds, if they are to lead to 
explicit consistent results, and will not be considered here. 
Instead, we utilize the linkage of the above action 
(and actions of this broad type) to Nambu Brackets (NB).

The classical variational equations of motion of the action resulting 
from $\delta z^i$ are
\be
0= (\epsilon^{ijk} \partial_t z^j +  \partial_i H \partial_k G -      
\partial_k H \partial_i G ) ~\partial_\sigma z^k .
\ee
Motion along the string  (along $\partial_\sigma z^i$) may be 
gauge-fixed by virtue of $\sigma$ reparameterization invariance,\cite{lund}
so that this reduces to
\be
\dot{z}^i =  \epsilon^{ijk} \partial_j H \partial_k G ~,
\ee
and these amount to Nambu's dynamical equations,\cite{nambu} 
\be
\dot{z}^i  = {\partial (z^i,H,G) \over \partial (z^1,z^2,z^3)   }\equiv 
\{ z^i, H,G \} . \label{motion}
\ee
In this context, this Jacobian determinant (volume element), defines 
the Nambu 3-bracket, which generalizes and supplants the Poisson Bracket, 
and is likewise linear and antisymmetric in its arguments ---here, with 
two ``Hamiltonians", instead of one. Thus, for an arbitrary function $f(z^i)$, 
\be
{df\over dt}= \{ f, H,G \} ~.  \label{3CNB}
\ee
As manifest from this, $H$ and $G$ are time-invariant, and the above velocity 
is divergenceless, $\partial_i \dot{z}^i=0$,  
(solenoidal flow, {\em cf.}\ the Appendix). 
In form language, the ``Cauchy characteristics"\cite{chern} are directly 
read off the 
3-form (\ref{3form}), whose first variation yields the above equations,
since 
\be 
d\omega_2 \! =\!  (dz^1\!  -\! \{ z^1 , H, G \}dt)\! \wedge\!  
(dz^2\! -\! \{ z^2 , H, G \}dt)\! \wedge\!  (dz^3\!  -\! \{ z^3 , H, G \}dt). 
\ee

The generalization of the above 3-form 
illustration to an arbitrary exact $p$-form, 
$d\omega_{p-1}=dz^1\wedge ... \wedge dz^p - dt\wedge 
dI_1\wedge ... \wedge dI_{p-1}$, hence a $(p-2)$-brane, is 
straightforward.\cite{estabrook,takhtajan,matsuo,pioline} The 
$I_i$s  are invariants (``Hamiltonians") entering into $p$-NBs.\cite{talks}
(Formally, it may describe $(p-2)$-branes moving 
in $p$-dimensional spacetimes; 
$p=2$ reduces to Hamiltonian particle mechanics in phase space 
and Poisson Brackets.)   For such topological systems,\cite{talks} 
``open membranes" is a bit of a misnomer, only adhered to 
for historical reasons. 
They are akin to D-branes, as they represent the dynamics of sets of 
points $z^i$ which do not really influence the motion of each other:
the membrane coordinates (string coordinate $\sigma$ above) 
are only implicit in the  $z^i$s; and their number may only be inferred 
from the above action whose formulation they expedited---but they do 
not enter explicitly in Nambu's equation of motion,
\be
{df\over dt}=  \{f,I_1,...,I_{p-1}\} ~.\label{pCNB} 
\ee

Now, consider introducing another variable, $z^4$, in 
the 3-NB (\ref{3CNB}), to convert it to a 4-NB,
\be 
{df\over dt}=  \{f,H,G\}= 
{\partial (f,H,G,z^4) \over \partial (z^1,z^2,z^3,z^4)   }\equiv 
\{f,H,G,z^4\}.\label{pplusCNB} 
\ee
In general, one may thus promote any $p$-NB to a $(p+1)$-NB,\cite{mukunda}
but we will focus on the case of {\em odd} $p$ upgrading to an {\em even} 
$p+1$. 

The reason is that even-$p$-NBs are ``nicer".\cite{CQNB,Azcarraga}
One may think of the $p=2n$ variables $z^i$ as paired phase-space 
variables $q^i, p^i$ of a system of $n$ degrees of freedom. The 
$2n$-NBs always resolve to a fully antisymmetrized sum of strings of 
PBs in that phase space, of all pairs of the NB 
arguments.\cite{CQNB,sphere,talks} For example, 
for a 4-NB,
\be 
\{  I_1,I_2,I_3,I_4\}  
=\{ I_1 ,I_2 \} \{  I_3 ,I_4 \}- \{ I_1 ,I_3\} \{  I_2 ,I_4 \}  
-\{  I_1 ,I_4 \} \{  I_3 ,I_2 \} ,\quad \label{resolution} 
\ee
($=\epsilon^{ijkl} \{ I_i,I_j\} \{ I_k,I_l\}  /8$. 
K Bering has observed that, in general, the PB resolution of
the  $p=2n$-NB amounts to the Pfaffian of the 
antisymmetric matrix with elements $\{ I_i,I_j \}$.)

Perhaps equally importantly, $2n$-NBs automatically describe 
classical motions in phase space for all Hamiltonian maximally 
superintegrable systems with $n$ degrees of freedom,
{\em i.e.}, systems with $2n-1$ algebraically independent integrals of 
motion.\cite{nutku,sphere,CQNB} Here, 
motion is confined in phase space on the constant surfaces 
specified by these integrals, and thus their collective intersection: 
so that the phase-space velocity 
${\bf v}=(\dot{{\bf q}},\dot{{\bf p}})$ is always perpendicular to the
$2n$-dimensional phase-space gradients 
$\nabla=(\partial_{\bf q}, \partial_{\bf p})$ of these integrals of the motion. 
Consequently, the phase-space velocity
must be proportional to the generalized cross-product of all those gradients,
and for any phase-space 
function $f({\bf q},{\bf p})$, motion is fully specified by the NB of eqn 
(\ref{pCNB}), with a prefactor $V$
\be 
{df\over dt}=\nabla f\cdot {\bf v} = V ~\partial_{i_1} f ~ 
\epsilon^{i_1i_2...i_{2n}} ~ 
\partial_{i_2} I_{1} ... \partial_{i_{2n}}I_{2n-1} ~.  \label{cross}
\ee
The proportionality constant $V$ is known to be 
time-invariant\cite{sphere,CQNB} and is a function of the invariants $I_i$,
so that $\nabla \cdot{\bf v}=0$ (solenoidal flow). 

Thus, there is an abundance of simple classical symmetric systems controlled 
by NBs,
such as multioscillator systems, chiral models, or free motion on 
spheres, even if they are also describable by Hamiltonian 
dynamics.\cite{sphere,CQNB,nutku,chatterjee} Through the embedding 
conversion of the type
(\ref{pplusCNB}), it is then evident that this extends to odd-NBs as well.

\begin{figure}[ht]
\centerline{\epsfxsize=6.5in\epsfbox{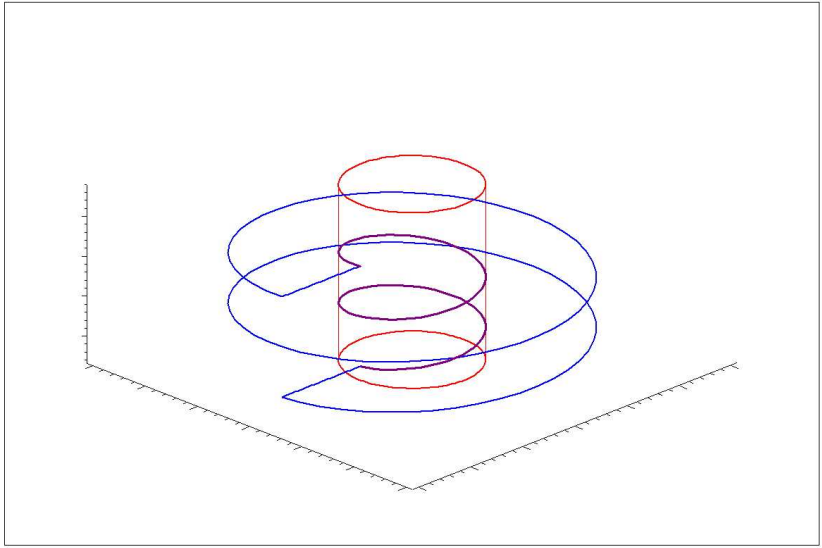}}   
\caption{A cylinder of constant $H$ intersects a helical ramp of
constant Re $G$ to define a trajectory.\label{archimedesfig}}
\end{figure}

As an illustration of (\ref{pplusCNB}), consider a simple system
in this phase-space language, ($z^1=x,~z^2=p_x,~z^3=y,~z^4=p_y$), 
{\em cf.}\ Fig~\ref{archimedesfig}, 
\be
H={p_x^2+x^2 \over 2} ~, \qquad G=i\ln (p_x+ix) + y ~. \label{triv}
\ee

Since a determinant is unaffected under
combination of its columns, the 4-NB in eqn (\ref{pplusCNB}) 
is the same when the extraneous coordinate $p_y$ is introduced into a 
modification,
$\tilde{H} \equiv H+p_y$, (so that  $\{\tilde{H} ,G\}=0=\{\tilde{H} ,p_y\}$.) 
Then, by eqn ({\ref{resolution}),
\be
{df\over dt}=  \{ f, \tilde{H} \}, \label{neatevol}
\ee
and so yields a superintegrable Hamiltonian system.
 
By contrast, suppose one took instead, 
\be
\tilde{H} ={p_x^2+x^2 \over 2}+p_y ~, \quad 
\tilde{G} =(p_x+ix)\exp (-iy) ~, \label{freak} 
\ee
hence 
\be
{df\over dt}=
 -i\tilde{G} \{ f,\tilde{H}  \} =-i\{\tilde{G} f, \tilde{H}\}
,\label{archimedeanevol}
\ee
which is {\em not} a Hamiltonian system. For a Hamiltonian system, the 
four Hamilton's equations have 6 cross-consistency conditions 
among them. For this system, though, even though $\nabla \cdot{\bf v}=0$,
nevertheless
\be
\partial_x \dot{x}+ \partial_{p_x} \dot{p}_x- \partial_{y} \dot{y}- 
\partial_{p_y} \dot{p}_y= 2\tilde{G} \neq 0,  
\ee
in violation of Hamiltonian consistency. The trajectories in phase space 
are the same as above, but the speed along 
each trajectory depends on the value of $\tilde{G}$ on that trajectory.

The evolution  (\ref{archimedeanevol}) in this auxiliary phase space 
specifies a non-flat Poisson structure,
${\stackrel{\leftarrow}{\partial}_i}\alpha^{ij}{\stackrel{\rightarrow}
{\partial}_j}$ (which satisfies the Jacobi identity,\cite{nutku} 
hence $d(\alpha^{-1})=0$). 
Such Poisson manifolds are defined {\em in general for all}\cite{nutku} 
{\em NB evolution laws} (\ref{cross}). 
Like the curved Poisson manifold evolutions of 
constrained Hamiltonian systems in $2p$-dim phase space discussed earlier, 
these flows could also be quantized in phase space, in principle, by 
Kontsevich's\cite{Kontsevich} elaborate deformation algorithm leading to 
Moyal brackets (which also satisfy the Jacobi identity). In general, 
it is not evident that such results do not differ from those of the methods 
considered here. 

We will consider in the next section quantization of NB evolutions 
simplifying  through PB resolutions into 
\be
{df\over dt}= V \{f, I_1, ... ,I_{2n-1}  \}   
=  W \{ f, H \}, \label{classicalevolution} 
\ee
for some functions $H(I_i), V(I_i), W(I_i)$.  $W$=constant  
({\em e.g.}\ eqn (\ref{neatevol})) amounts  to 
maximally superintegrable Hamiltonian systems, quantized and illustrated 
extensively in Refs \refcite{sphere,CQNB,talks}. 

 Classically, by linearity in all derivatives involved, 
all three members of this equation are derivations, {\em i.e.}~satisfy 
Leibniz's rule, ~~$\delta (fg)= \delta (f) ~g + f~\delta(g)$. 

By contrast, eqn (\ref{archimedeanevol}) has $V=1, W= -i\tilde{G}$.
To be sure, $W\neq$const systems might well be convertible
to hamiltonian ones, by virtue of non-canonical 
transformations---necessarily,
if the hamiltonian structure is {\em not} to be preserved. {\em E.g.}, for the 
system (\ref{freak}), the four equations of motion 
resulting from (\ref{archimedeanevol}) 
can be rewritten through the change of variables, 
\be
p_w \equiv\tilde{G} , \qquad w \equiv iy ~ \frac{p_x^2 +x^2}{ 2\tilde{G}} ~,
\ee
to
\be
      \dot{x} =  p_x p_w ~, \qquad 
      \dot{p}_x= -x p_w ~,   \qquad 
      \dot{p}_w =0 ~, \qquad 
      \dot{w} = \frac{p_x^2 +x^2}{2}  ~,
\ee
which are derivable from a Hamiltonian, 
$~{\mathcal H}= p_w (p_x^2  +x^2)/2$.

\section{Quantum Nambu Dynamics}\label{sec:quantum} 

To quantize eqn (\ref{classicalevolution}), we seek a map to 
operators in Hilbert space corresponding conventionally to the classical 
quantities, which preserves as much of the structure and the symmetries of it
as possible,\cite{CQNB} or, equivalently, a deformation in phase space 
relying on a $\star$-product which acts on the same phase-space 
functions.\cite{sphere} (When $W$=constant, the answer must 
coincide with results provided by Hamiltonian quantization.) 
We have found that several consistency complications are not debilitating.

Let us adopt Nambu's proposal\cite{nambu} of
a fully antisymmetric multilinear generalization of Heisenberg commutators,
now called quantum NBs (QNB):
\be
\left [ A, B \right ] \equiv AB-BA , 
\ee
\be
\left[ A,B,C\right] \equiv    A B C-A CB+BCA-BAC+CAB-CBA,  \label{3qnb}
\ee
\be 
\left[ A,B,C,D\right] \equiv   A [ B,C,D ] -B \left[ C,D,A\right] 
+C\left[D,A,B\right] -D \left[ A,B,C\right]   \qquad \qquad \label{4qnb}
\ee
$$
=  [A,B][C,D]+ [A,C][D,B]+ [A,D][B,C]
$$
$$
+[C,D][A,B]+ [D,B][A,C]+ [B,C] [A,D]~,
$$  
{\em etc.} As for classical NBs, all even-QNBs resolve into strings of 
commutators,\cite{CQNB} with all suitable inequivalent orderings required 
for full antisymmetry, ($\propto \epsilon^{ijkl...} 
[ I_i,I_j] ~[I_k,I_l] ...$): a ``quantum Pfaffian". 
Consequently, they manifestly have the proper classical 
limit\cite{CQNB} to the classical NBs of the previous section, 
\be
[I_1,...,I_{2n}] ~~\rightarrow ~~ n! ~(i\hbar)^n ~\{I_1,...,I_{2n}\}.
\ee 

By contrast, odd-QNBs have a bad classical limit\footnote{This is seen 
most directly in phase-space quantization, as 
$\hbar \rightarrow 0$, since the $\star$-products intercalated 
among all functions in, {\em e.g.}, (\ref{3qnb}), yield an even number of 
derivatives---not an odd one, as required by the classical  NBs
such as eqn (\ref{motion}). }, underlying our 
conversion to even-NBs at the classical level, (\ref{pplusCNB}). 
 Thus, {\em e.g.}, taking $f=p_y$ and eqn (\ref{triv}),
it is easy to see that a naive quantum counterpart of 
(\ref{pplusCNB}) fails, 
\be
[f,H,G,p_y]\neq \text{const } [f,H,G]. \label{4neq3}
\ee

It is the 4-QNB on the left-hand side 
which we choose for quantization below, and its even generalizations 
throughout. Even though the extraneous variable $p_y$ does not 
enter in the description of the classical trajectories, its operator
counterpart is unavoidable in Hilbert space, as it is the commutator conjugate 
to the operator $y$.
 
A further complication of  odd-QNBs\cite{nambu} 
is also remedied by the odd-to-even embedding adopted.
In quantizing (\ref{classicalevolution}), if the dynamical variable
(whose time evolution involves its entering in a QNB) is taken to 
be the identity operator 
(e.g.~$f\propto {1\kern-0.36em\llap~1}$), its time derivative vanishes. 
Whereas, {\em e.g.}~for eqn (\ref{3qnb}),
\be
[ {1\kern-0.36em\llap~1}, B,C] = [B,C]\neq 0~, \label{identitycrisis}
\ee
which restricts the specification of $df/dt$ through odd QNBs, in 
general\footnote{For privileged cases, {\em e.g.}~$[B,C]=0$, there 
is no such problem, {\em cf.}~the Appendix.}.
 
In contrast, this {\em does} result in the vanishing of all even-QNBs,
identically, such as (\ref{4qnb}), by virtue of their commutator resolution,
\be
[ {1\kern-0.36em\llap~1}, I_1,...,I_{2n-1}] = 0~. \label{evennocrisis}
\ee

For other aspects of this odd-even dimorphism see Refs 
\refcite{Hanlon,Azcarraga,CQNB}.

The associative strings of operators making up 
the $2n$-QNBs, satisfy the proper fully antisymmetric 
Generalized Jacobi Identity.\cite{Hanlon,Azcarraga,CQNB}
{\em E.g.}, for 4-QNBs, (\ref{4qnb}),
\be
\epsilon^{ijklmnr} [[I_i ,I_j,I_k, I_l],I_m,I_n,I_r] 
=0.
\ee

However, the above $2n$-QNBs forfeit Leibniz's derivation property 
(beyond the quantum commutator, the 2-QNB), in general. As a result,
they do not satisfy the classical identities arising from the 
derivation property (sometimes referred to as ``fundamental 
identities";\cite{Sahoo,takhtajan} for the above 4-QNBs, those would have 
5, instead of the above $7!/4!3!=35$ terms.) 
This may only be a subjective shortcoming, dependent on 
the specific application context: if a choice is forced, associativity
trumps the derivation property.\cite{Azcarraga}

For example, in several Hamiltonian maximally superintegrable physical 
systems,\cite{sphere,CQNB,talks} with  
$W$=constant, and arbitrary  $V$ in eqn (\ref{classicalevolution}),  
quantization was demonstrated explicitly to be consistent. 
The QNBs failure to be a 
derivation is in accord with its being equal to operator 
expressions corresponding to the classical quantities 
\be
{1 \over V} {df\over dt} ~. 
\ee
Upon quantization, such expressions involve entwinements of derivatives 
of operators $df/dt$ 
with other operators, such as 
\be
{1 \over V} {df\over dt} + {df\over dt} {1 \over V}~~,
\ee
but often with longer, more involved strings of operators.\cite{CQNB}
Such strings also fail to be derivations, although they can be demonstrated to 
provide consistent results equivalent to Hamiltonian quantization.
The formal Jordan-Kurosh spectral problem\cite{Kurosh} involving 
such long associative operator strings may be involved technically,
but yields consistent answers.\cite{CQNB} 

Of course, in exceptional cases (such as those with $W$=const, $V$=const), 
$2n$-QNBs can be derivations, if only because they are equivalent to quantum 
commutators. For example, quantization of eqn (\ref{neatevol}) by (\ref{4neq3}) 
trivially yields, by the commutator resolution (\ref{4qnb}),
\be
{df\over dt}=-{1\over 2 \hbar^2} [f,H,G,p_y]={ 1\over i \hbar} [ f, 
\tilde{H} ], \label{zellweng}
\ee
where the operator $p_y$ corresponding to the extraneous variable is now 
featured in the effective hamiltonian. (The eigenstates of 
$\tilde{H}$ are thus $\psi(x,y,n,k)\propto H_n(x/\sqrt{\hbar}) ~
\exp (iky-x^2/\hbar)$.)\footnote{
To be sure, not all classical features of Nambu dynamics transfer over to 
the quantum domain without complication. A more technical treatment of 
departures from formal quantum operator solenoidal flow is provided in 
the Appendix.} 

Our conclusion, then,  is that the embedding of odd to even brackets 
of rank one higher enables the corresponding membranes to rely on the 
consistent results on even-QNBs, as described, for their quantization. 
The toolbox provided in Ref \refcite{CQNB} may guide quantization of 
more general systems beyond Hamiltonian ones, {\em i.e.}\ systems 
with nontrivial $W$. A further important question is whether 
the results (such as spectra) of the QNB quantization discussed here 
coincide with those obtained from simple systems quantized through Kontsevich's
product---but in those cases explicit answers are normally hard to come by.

\section*{Acknowledgments}
We thank the organizers of this conference for their hard work and 
hospitality, and also Y Nambu, D Fairlie, Y Nutku, J de Azc\'{a}rraga, 
A Stern, A Polychronakos, K Bering, V S Nair, and V Gueoruiev for insights.

This work was supported by the US Department of Energy, 
Division of High Energy Physics, Contract W-31-109-ENG-38,
and the NSF Award 0303550.

\appendix

\section{Solenoidal Flow}

\renewcommand{\theequation}{A.\arabic{equation}}

Preservation of phase-space volume through solenoidal flow, 
$\nabla \cdot {\bf v}=0$, (Liouville's theorem) was a guiding
principle for Nambu,\cite{nambu}  who confirmed it for 
the classical dynamics of his original {\em maximal} NB. We show 
that all flows generated by NBs, even submaximal ones, are solenoidal, 
but, in general, the ones formally generated by QNBs are not. 

For even classical maximal brackets in $2n$-dimensional phase space,
it was made evident for (\ref{cross}) that $\nabla \cdot {\bf v}=0$, 
since 
\be 
v^{i_1}=\left\{ z^{i_1},I_{1},\cdots ,I_{2n-1}\right\} =
\epsilon^{ i_{1}\cdots i_{2n}}~\partial _{i_{2}} I_{1}\cdots 
\partial _{i_{2n}} I_{2n-1}.
\ee
Even sub-maximal NBs are obtained from maximal ones 
for $2k-1$ fixed $I$s with $k<n$, through symplectic tracing of maximal 
brackets,\cite{CQNB}
\be
&\phantom{=}& \left\{  z^{j},I_{1},\cdots ,I_{2k-1}\right\}  \\
& =& \frac{1}{\left(
n-k\right) !}\sum_{\text{all }i=1}^{n}\left\{ z^j ,I_1,\cdots
,I_{2k-1},x_{i_{1}},p_{i_{1},}\cdots ,x_{i_{n-k}},p_{i_{n-k}}\right\} 
\nonumber .
\ee
Hence, since each term in the sum produces a solenoidal flow,
the sum of terms is also solenoidal,
${\partial_j } ~\left\{ z^j,I_{1},\cdots ,I_{2k-1}\right\}=0$.

Odd-NBs are obtainable through the embedding conversion 
described, and can likewise be made submaximal, hence 
they generate solenoidal flows as well,
\be 
0 &=&\partial_j \left\{ z^j,I_{1},\cdots
,I_{2k-2}\right\} \\
&\equiv &\frac{1}{\left( n-k\right) !}\sum_{\text{all } i=1}^{n} \partial_j 
\left\{ z^j,I_{1},\cdots
,I_{2k-2},x_{i_{1}},p_{i_{1},}\cdots ,x_{i_{n-k}},p_{i_{n-k}},p_{2n}\right\}
\nonumber .
\ee

The direct quantum operator analog of the above classical 
$\nabla \cdot {\bf v}$ (reducing to it in the classical limit) is 
\be
K(I_i,...)\equiv J_{ij} [z^i, [ z^j,I_1,I_2,\cdots ] ],
\ee
where $J_{ij}$ is the standard symplectic metric for the canonical variables, 
$[z^i, z^j ] =i\hbar J_{ij}$. In general, $K$ does {\em not} vanish, 
even though in special circumstances, such as (\ref{freak}), it can vanish.
 That is, 
{\em e.g.}~for 3-QNBs,
\be 
K= J_{ij } [ z^i, [ z^j ,H,G ] ] =\frac{1}{2}J_{ij} ~ 
 [ z^i ,z^j ,H,G ] -J_{ij} z^i [ z^j,[ H,G ] ] ~ .
\ee 
For $\tilde{H} $ and $\tilde{G}$, both their commutator and 
the quantum symplectic trace 
($J_{ij} [ z^i, z^j ,\tilde{H},\tilde{G} ] ] =0$) vanish.

In general, however, even for privileged flows 
for which $\left[ H,G \right] =0$  (so that they 
leave the unit operator invariant, by eqn (\ref{identitycrisis}): ~  
$[ {1\kern-0.36em\llap~1}, H,G] = [H,G]=0$),
the above symplectic trace is non-vanishing, as can be seen by 
taking brackets of exponentials linear in the canonical variables:
\be
&J_{lj}& 
\Bigl  [ z^l, z^j, \exp \left( i\alpha \cdot z\right) ,\exp \left(
i\beta \cdot z\right) \Bigr ]  \\
& =&i\hbar \left( 4n-\frac{2\hbar \alpha\wedge \beta }
{\tan \left( \frac{\hslash }{2}\alpha \wedge \beta \right) } \right) 
~ \Bigl    [ \exp \left( i\alpha \cdot z\right) ,
\exp \left( i\beta \cdot z\right) \Bigr  ]~, \nonumber
\ee
where $\alpha \wedge \beta \equiv \alpha^i J_{ij}\beta^j$. 
Operator Fourier analysis of general $H$ and $G$
evinces the quantum symplectic trace to not be necessarily proportional
to $[ H,G]$. 

In general, $K$ reduces to
the antisymmetric {\em derivator}\cite{CQNB} for canonical variables. 
 The derivator for any given string of $A$s is defined for all $B$ and $C$ as 
\be 
( &B&,C\ |\ A_{1},\cdots ,A_{k} ) \\
&\equiv&\left[ BC,A_{1},\cdots ,A_{k} \right] 
-B\left[ C,A_{1},\cdots ,A_{k}\right] -\left[ B,A_{1},\cdots ,A_{k}\right] C,
\nonumber 
\ee 
so that it vanishes for \emph{all} $B$ and $C$ iff the 
$(k+1)$-QNB-generated action of the string of $A$s is a derivation.

A sufficient condition for formal ``operator solenoidal flow" 
follows from the identity,
\be 
&J_{ij}& \left[ z^i ,\left[ z^j ,A_{1},\cdots ,A_{k}\right] \right]\\
 &=& i\hbar n ~ [ {1\kern-0.36em\llap~1}  ,A_{1},\cdots ,A_{k} ] 
-J_{ij}\ \left( z^i,z^j\ |\ A_{1},\cdots ,A_{k}\right). \nonumber  
\ee
For even $k+1\equiv 2m$, by eqn (\ref{evennocrisis}), 
the first term on the r.h.s.\ drops out, and 
$K$ amounts to minus the derivator. 

For odd QNBs, even $k=2m$, as for (\ref{identitycrisis}), 
$\left[ {1\kern-0.36em\llap~1},A_{1},\cdots ,A_{2m}\right] 
=\left[ A_{1},\cdots ,A_{2m}\right] $ may 
not vanish in general, but one might consider the 
privileged ``flows" for which it does. Thus, for such privileged 
odd-QNB-generated  operator flows and for all 
even-QNB-generated flows, 
\be
K=J_{ij}\ \left[ z^i ,\left[ z^j ,A_{1},\cdots ,A_{k}\right] \right]
=-J_{ij}~\left( z^i,z^j\ |\ A_{1},\cdots ,A_{k}\right) .
\ee
Quantum flows are thus solenoidal if the action of the 
$A_{1},\cdots ,A_{k}$ in an $(k+1)$-QNB is a derivation. 

However, it is not necessary for the general derivator to vanish to
have operator solenoidal flow---just this particular derivator. For the 
example considered above, $\tilde{H}$, $\tilde{G}$, a 3-QNB 
based on them is not a derivation, in general: \ 
$\left( B,C\ |\ \tilde{H},\tilde{G}\right) \neq 0$ for all $B$ and $C$. 
Nonetheless, as noted, $K$ vanishes for this case, as well as for the 
3-QNB system with $\tilde{H}, G$, eqs (\ref{triv}, \ref{freak}), 
 and the 4-QNB system with $\tilde{H}, \tilde{G}, p_y$, eqs (\ref{triv}, 
\ref{zellweng}). 
%
%
%
%

\end{document}